\documentclass[11pt]{article}
\usepackage{latexsym}
\usepackage{amssymb}
\usepackage{amsmath}
\usepackage{graphicx}
\usepackage{subfigure} 
\usepackage{amsfonts}
\usepackage{array}

\numberwithin{equation}{section}

\newcommand{\be}{\begin{equation}}
\newcommand{\ee}{\end{equation}}
\newcommand{\bea}{\begin{eqnarray}}
\newcommand{\eea}{\end{eqnarray}}

\begin{document}
\begin{titlepage}

\begin{center}
{\Large {\bf Spherically Symmetric Black Hole Formation in
   Painlev\'e-Gullstrand Coordinates}}

\vspace{0.125in}

{\bf
Jonathan Ziprick${}^\dagger$
and
Gabor Kunstatter${}^\sharp$
}

\vspace{0.25in}

{\sl
${}^\dagger$ 
Department of Physics and Astronomy\\
University of Manitoba\\
Winnipeg, Manitoba, Canada R3T 2N2\\
{[e-mail: j.ziprick-ra@uwinnipeg.ca]}\\
}

\vspace{0.125in}

{\sl
${}^\sharp$ 
Department of Physics and Winnipeg Institute of Theoretical Physics\\ 
University of Winnipeg\\
Winnipeg, Manitoba, Canada R3B 2E9\\
{[e-mail: g.kunstatter@uwinnipeg.ca]}\\
}

\vspace{0.5in}

\begin{abstract}
We perform a numerical study of black hole formation from the spherically symmetric collapse of a massless scalar field. The calculations are done in Painlev\'e-Gullstrand (PG) coordinates that extend across apparent horizons and allow the numerical evolution to proceed until the onset of singularity formation. We generate spacetime maps of the collapse and illustrate the evolution of apparent horizons and trapping surfaces for various initial data. We also study the critical behaviour and find the expected Choptuik scaling with universal values for the critical exponent and echoing period consistent with the accepted values of $\gamma=0.374$ and $\Delta = 3.44$, respectively. The subcritical curvature scaling exhibits the expected oscillatory behaviour but the form of the periodic oscillations in the supercritical mass scaling relation, while universal with respect to initial PG data,  is non-standard: it is non-sinusoidal with large amplitude cusps.
\end{abstract}

\end{center}

\end{titlepage}

\section{Introduction}
Black holes are among the most interesting solutions to Einstein's field equations. They demonstrate in various ways the incompleteness of our classical understanding of gravity and provide hints toward developing a quantum theory. The relevant issues are the endpoint of Hawking radiation, the breakdown of general relativity at curvature singularities and the microscopic source of black hole entropy. Given the lack of experimental guidance, black holes provide a vital testing ground for new ideas in quantum gravity. 

Despite a great deal of analytic and numerical work, relatively little is known about the dynamics of black hole formation except in special, highly symmetric situations, since the equations for collapsing matter fields tend to be very difficult to work with. 
%Much of our present knowledge is based on stationary vacuum solutions or the addition of matter as a small perturbation in a black hole spacetime. 
Important pioneering work in this area was done by Choptuik \cite{Choptuik} in the early 1990's, who took advantage of modern computational methods to numerically create small mass black holes from the collapse of spherically symmetric massless scalar fields. In the small mass limit he observed unexpected critical behaviour that has since been confirmed in numerous papers for a variety of matter fields (see \cite{Gundlach2} for an in-depth review).

There are three fundamental properties of the critical collapse that are relevant here. First, there is a critical zero mass black hole solution that plays the role of an intermediate attractor for the dynamical evolution. In the case of massless scalar fields the critical solution exhibits discrete self similarity (DSS) described in terms of a generic set of field variables by
\begin{equation}
\label{DSS}
Z_*(r,t_s) = Z_*(e^{m\Delta}r, e^{m\Delta}t_s),
\end{equation}
where $Z_*$ is the critical solution, $r$ and $t_s$ are the Schwarzshild radius and time, $m$ is some integer $\ge 0$ and $\Delta$ is a dimensionless number. To explain this more clearly, consider the profiles of the critical solution field variables frozen in time at $t_{s_0}$. After a certain amount of time $\delta t_s$ elapses, the same profiles will be found on a scale that is smaller by a factor of $e^\Delta$. After an additional time $\delta t_s/e^\Delta$ has passed, the same profile will again be found on a scale that is $e^{2\Delta}$ times smaller than at $t_{s_0}$. This gives rise to the more descriptive name for DSS in gravitational collapse, scale echoing.

The existence of this intermediate attractor gives rise to scaling relationships for near critical solutions. For any one parameter family of data in the limit of small black holes one finds a mass scaling relation:
\begin{equation}
\label{mass scaling}
\ln{(M_{BH})} \sim \gamma \ln{|p-p_*|} + f(ln|p-p_*|),
\end{equation}
where $M_{BH}$ is the black hole mass on formation, $\gamma$ is a real parameter referred to as the critical exponent, $f(ln|p-p_*|)$ is a periodic function, $p$ is any parameter in the initial data and $p_*$ is the threshold value of that parameter that separates the data into two categories: supercritical (leading to black hole formation) and subcritical (leading to matter dispersion). Two separate analytical studies (\cite{Gundlach1}, \cite{HP}) have shown that DSS results in the period of $f(ln|p-p_*|)$ being
\begin{equation}
\label{period}
T = \Delta/2\gamma.
\end{equation}
Note that $f(ln|p-p_*|)$ is often referred to as a ``small oscillation" or ``wiggle", but its explicit functional form must be determined numerically \cite{Gundlach1}.

Garfinkle and Duncan have shown that a similar scaling relation exists for the maximum curvature at the origin for subcritical solutions \cite{GD}. For any one parameter family of data in the barely subcritical limit one finds:
\begin{equation}
\label{R scaling}
\ln{ \left(\hbox{MAX}_t \left\{ R(r=0) \right\} \right)} \sim -2 \gamma \ln{|p-p_*|} + g(ln|p-p_*|),
\end{equation}
where $R(r=0)$ is the Ricci scalar at the origin and $g(ln|p-p_*|)$ is a periodic component, again with period $T$. 
%$s$ is the proper time of a central observer given by
%\begin{equation}
%\label{proper time}
%s \equiv \int_0^s \sqrt{g_{00}(0,t)}\;dt.
%\end{equation}
This relation is similar to that of mass scaling except for the factor in front of $\ln{|p-p_*|}$ being $-2 \gamma$ rather than $\gamma$. This derives from mass having units of length and curvature having units of length${}^{-2}$ (with $c=\hbar=1$).

The third aspect of critical collapse we note here is universality. DSS and critical scaling have been observed in the gravitational collapse of many different types of fields. The constants $\gamma$ and $\Delta$ are determined by the properties of the critical solution and are therefore universal: they vary for different kinds of matter but are independent of the form of initial data and the parameter $p$ in that initial data that is varied. Numerical and semi-analytical calculations for the massless scalar field have given $\gamma \simeq 0.37$ and $\Delta \simeq 3.4$ (\cite{Gundlach1}, \cite{HP}, \cite{Choptuik}, \cite{Garfinkle}, \cite{HS}, \cite{Bland}). In all previous calculations, done in Schwarzschild or double null coordinates, the functions $f$ and $g$ were well approximated by a small amplitude sine function with period $T = \Delta/2\gamma \simeq 4.6$.

In this paper we study numerically black hole formation from the spherically symmetric collapse of a massless scalar field in Painlev\'e-Gullstrand coordinates (\cite{Painleve}, \cite{Gullstrand})\footnote{These coordinates are sometimes referred to as ``flat slice'' coordinates because in vacuum the surfaces of constant time are flat.}. PG coordinates were used by Wilczek and collaborators \cite{Wilczek} to study Hawking radiation and more recently were proposed by Husain and Winkler \cite{Husain05} as a reasonable starting point to investigate the quantum dynamics of black holes. The key distinguishing feature of PG coordinates is that they are regular across apparent horizons. This allows us to evolve initial data until the onset of singularity formation when our chosen boundary conditions (zero mass density at the origin) are no longer valid.  

We first of all take advantage of the PG coordinate system to run the code beyond initial horizon formation to study in detail the formation of apparent horizons and the evolution of the resulting trapping surfaces.  We then study the critical behaviour of the system and find our data to be in good agreement with previous results for the critical exponent and echoing period. Interestingly, we find the maximum curvature scaling relation to show the usual small oscillations but discover a new form for the oscillations in the mass scaling relation. The form of these mass scaling oscillations is found to be universal in the sense that it is the same for all families of initial data (that we tested) specified in PG coordinates. They are however quite different from those observed in Schwarzschild or double null coordinates. The reason for this difference is as yet unclear, but it is almost certainly related to the fact that the location of the initial apparent horizon can depend on the choice of spacetime slicings.

%The reason for this difference is not as yet fully understood (by us), but is almost certainly related to the form of the critical solution in PG coordinates.

While these results are interesting in themselves, our analysis is also important because it paves the way for studying the effects of quantum corrections on both the critical behaviour and the dynamics of trapping surfaces \cite{KZ}.

The paper is organized as follows: in Section \ref{sec: eqmo} we lay out the foundation of our mathematics and develop the equations of motion used in our computer code. Section \ref{sec: comp meth} describes our code giving relevant details of the numerical techniques and boundary conditions. Following this, we present the results of our simulations, discussing the scaling relationships and global properties of the spacetime (including trapping surfaces). We conclude with some remarks on these findings and consider future directions for this work.

\section{Equations of Motion}
\label{sec: eqmo}

The dynamical system that we wish to study is that of a collapsing spherically symmetric, massless scalar field in four spacetime dimensions. It turns out to be convenient to express these equations in a somewhat non-traditional parameterization, using the formalism of 2d dilaton gravity (see \cite{GKV} for a general review of dilaton gravity and \cite{KPS} for applications to black holes). In addition to providing a relatively simple form for the evolution equations in 4 dimensions, the formalism is generally applicable to a variety of spherically symmetric theories in four or more dimensions \cite{Bland}. We will describe the formalism in some detail because the gauge fixing is a crucial part of our analysis. 
We begin by stating the classical action and defining a metric. We then perform a canonical transformation and fix the gauge to yield dynamical equations for the scalar field in PG coordinates. 
Throughout this paper we use a prime to represent $\partial_r$ and an overdot for $\partial_t$.

The action for a massless scalar field $\psi$ coupled to a dilaton in 2d is
\begin{equation}
S[g, \phi, \psi] = \frac{1}{2} \int dx dt \sqrt{-g}\left[\frac{1}{G}\left(\phi R(g) + \dfrac{V(\phi)}{l^2}\right) - h(\phi) \left|\nabla \psi \right|^2\right]
\label{eq:dilaton action}
\end{equation}
where $G$ is Newton's consant in 2d, $g$ is the metric determinant, $R(g)$ is the Ricci scalar and $h(\phi)$ is the dilaton coupling.

The relationship of this action to spherically symmetric gravity in $D=n+2$ spacetime dimensions is given as follows. The physical $D$ metric is:
\be 
ds^2_\text{phys}= \frac{ds^2}{j(\phi )} + r^2(\phi) d\Omega^2_n,
\label{eq:physical metric}
\ee
where $d\Omega^2_n$ is the line element of the unit $n$-sphere and
\begin{equation}
j(\phi):=\int_0^\phi d\tilde{\phi}V(\tilde{\phi}).
\end{equation}
With the further substitutions
\bea
\label{eq: G}
2G&=& \frac{16\pi G^{(n+2)} n}{8(n-1){\cal \nu}^{(n)}l^n},\\
\phi&=&\frac{n}{8(n-1)}\left(\frac{r}{l}\right)^n,\\
V(\phi)&=& (n-1)\left(\frac{n}{8(n-1)}\right)^{1/n}\phi^{-1/n},\\
h(\phi)&=& \frac{8(n-1)}{n}\phi=\left(\frac{r}{l}\right)^n,\\
\psi&=&\sqrt{{\cal \nu}^{(n)} l^n}\overline{\psi}, 
\label{eq:dictionary}
\eea
the dilaton action (\ref{eq:dilaton action}) is (up to boundary terms) precisely equal to that of a spherically symmetric massless scalar field $\overline{\psi}$ minimally coupled to Einstein gravity in $D$ spacetime dimensions:
\be
I^{(D)}= \frac{1}{16\pi G^{(D)}}\int d^Dx \sqrt{-g^{(D)}}R^{(D)}-\frac{1}{2}
   \int d^{D}x \sqrt{-g^{(D)}} |\nabla \overline{\psi}|^2.
\ee
In the context of spherically symmetric gravity, performing the dimensional reduction in the action before varying yields the same dynamics as imposing spherical symmetric in the field equations themselves.

In order to make our coordinate choice explicit and rigorous, we now summarize the Hamiltonian analysis for the theory. We first define the metric in modified ADM form:
\begin{equation}
\label{le1}
ds^2 := e^{2\rho}\left[-\sigma^2dt^2 + (dr+Ndt)^2\right]
\end{equation}
where $\rho$, $\sigma$ and $N$ are arbitrary functions of the spacetime coordinates.

The authors of \cite{DGK}  derive a partially gauge fixed Hamiltonian. For completeness, we review this procedure before applying a second gauge fixing condition that yields the completely gauge fixed metric in PG form. 

The first gauge choice is
\begin{equation}
\label{fix1}
j(\phi) = l \phi^\prime \,,
\end{equation}
which yields the consistency condition on the lapse and shift functions:
\begin{equation}
\label{sigma-N}
\sigma G \Pi_\rho = N \phi^\prime.
\end{equation}
The equations are put in a more transparent form using the canonical transformation:
\begin{eqnarray}
X&:=&e^\rho,\\
P&:=&e^{-\rho}G\Pi_\rho,
\end{eqnarray}
where $\Pi_\rho$ is the momentum conjugate to $\rho$. $P$ is then conjugate to $X$ with their Poisson bracket being
\begin{equation}
\left\{X(x),P(y)\right\}=G\delta(x,y).
\end{equation}
The resulting Hamiltonian is
\begin{equation}
\begin{split}
H(X,P,\psi,\Pi_\psi)=&\int dr \; \sigma \left(-\frac{X^2}{j(\phi)}{\cal M}^\prime + {\cal G}_{\cal M} + l\frac{XP\psi^\prime \Pi_\psi}{j(\phi)}\right)\\
                     &+\int dr \left(\frac{\sigma X^2}{j(\phi)}{\cal M} \right)^\prime
\end{split}
\end{equation}
where
\begin{eqnarray}
{\cal M} &=& \frac{l}{2G}\left(P^2-\frac{(\phi^\prime)^2}{X^2}+\frac{j(\phi)}{l^2}\right),\label{eq:cal M}\\
{\cal G}_{\cal M} &=& \frac{1}{2}\left(\frac{\Pi_\psi^2}{h(\phi)} + h(\phi)(\psi^\prime)^2\right).
\end{eqnarray}
${\cal M}$ is called the mass function because it approaches a constant at spatial infinity where it is equal to the ADM mass of the solution. ${\cal G}_{\cal M}$ is the energy density of the scalar field.

We now completely fix the gauge by choosing
\begin{equation}
\label{fix2}
X=\sqrt{j(\phi)}.
\end{equation}
This condition, along with Eqs.(\ref{eq:cal M}) and (\ref{fix1}), implies that $P^2 = 2G{\cal M}/l$. These gauge conditions produce the non-static generalization of PG coordinates, as can be seen by using (\ref{sigma-N}) and the gauge conditions in the case of a vacuum to reduce the line element to PG form \cite{KL}:
\begin{equation}
ds^2 = j(\phi)\left[-dt^2+\left(dr+\sqrt{\frac{2G{\cal M}l}{j(\phi)}}dt\right)^2\right].
\label{eq:vacuum PG}
\end{equation}
More importantly, the spatial slices will be seen to be regular across apparent horizons that form during the evolution.

We are now able to write the equations of motion for the scalar field in fully reduced form:
\begin{eqnarray}
\label{psi dot}
\dot{\psi} &=& \sigma \left(\frac{l \sqrt{2G{\cal M}/l} \psi^\prime}{\sqrt{j(\phi)}} + \frac{\Pi_\psi}{h(\phi)}\right),\\
\label{Pi dot}
\dot{\Pi}_\psi &=& \left[\sigma \left( h(\phi)\psi^\prime + \frac{l \sqrt{2G{\cal M}/l} \Pi_\psi}{\sqrt{j(\phi)}} \right) \right]^\prime,
\end{eqnarray}
where $\sigma$ and ${\cal M}$ are the solutions to
\begin{equation}
\label{P eq}
{\cal M}^\prime =  {\cal G}_{\cal M}+l\psi^\prime \Pi_\psi \sqrt{\frac{2G{\cal M}l}{j(\phi)}} ,
\end{equation}
\begin{equation}
\label{sigma eq}
\sigma^\prime + \frac{G l \psi^\prime\Pi_\psi }{\sqrt{2G{\cal M}l j(\phi)} } \sigma = 0.
\end{equation}
The right hand side of (\ref{P eq}) defines the instantaneous mass density of the configuration. It has the expected contribution from the energy density of the scalar and an additional coupling term that corresponds to the contribution from its self gravity. 
%For the spatial integrations, we begin with $P(0,t)~=~0$ and $\sigma(0,t)~=~1.0$ which implies a flat spacetime at the origin.
 
The above equations need to be supplemented by boundary conditions for $P$ and $\sigma$. Without loss of generality we choose $\sigma=1$ at $r=0$. A change in this value corresponds to a trivial rescaling of the time coordinate. We also fix ${\cal M}=0$ at $r=0$, which guarantees that the metric is flat in the neighbourhood of the origin. In fact, a non-zero value of ${\cal M}$ at the origin signals the formation of a singularity, so that this boundary condition would have to be relaxed/modified in order to integrate the equations past singularity formation.

\section{Computational Methods}
\label{sec: comp meth}

Since we work in four dimensions (n=2), we have $j(\phi)=\sqrt{\phi}$, $\phi=r^2/(4l^2)$ and $h(\phi)=4\phi$. For convenience we choose $2G=1$, which from Eq.(\ref{eq: G}) corresponds to choosing the characteristic length scale $l$ to be $l_{pl}=\sqrt{G^{(4)}}$. Note that the first gauge choice (\ref{fix1}) implies that our spatial coordinate is the radius $r$ of invariant two spheres.

The first step in the iteration process is to specify initial $\psi$ and $\Pi_\psi$ configurations. We work with two forms:
\begin{eqnarray}
\label{tanh}
\psi &=& A r^2 \exp{\left[-\left(\frac{r-r_0}{B}\right)^2\right]},\\
\label{gaussian}
\psi &=& A \tanh{\left(\frac{r-r_0}{B}\right)},
\end{eqnarray}
where $A$, $B$ and $r_0$ are the parameters which may be varied to study critical collapse. We shall refer to these two forms of initial data as gaussian and tanh respectively. Note that since the mass density depends on $\psi^\prime$, the tanh form has one mass peak while the gaussian data has two. For both cases we define an initial standing wave by choosing $\Pi_\psi(r,0)=0$.

The most interesting behaviour occurs as matter approaches the origin on ever finer scales. In order to observe this small $r$ behavoiur while keeping a large enough lattice to contain all of the mass within the system (numerical errors occur when mass leaves the grid), we refine the $r$-spacing $\Delta r(r)$ near the origin so that it is several orders of magnitude smaller than the spacing near the outer endpoint. We use an adaptive time step $\Delta t(t)$ refinement according to the minimum found across the spacial slice using the condition
\begin{equation}
\Delta t(t) = \hbox{MIN}_r\left\{\frac{dt}{dr} \Delta r(r)\right\}
\end{equation}
where $\frac{dt}{dr}$ is the inverse of the local speed of an ingoing null geodesic. This provides stability by preventing information from moving over too many $r$-points in a single time step. Having $\Delta t$ proportional to $\Delta r$ greatly increases computational times for finer $r$-spacing, placing a practical lower limit on the size of black holes that we are able to create. However, this is not a significant concern as we are able to work on small enough scales to produce critical mass scaling.

Both time and space integrations are performed using Runge-Kutta methods accurate to fourth order in the local grid spacing. 
%For the spacial integrations, we begin with $P(0,t)~=~0$ and $\sigma(0,t)~=~1.0$ which implies a flat spacetime at the origin. 
For derivatives, we experimented with finite differences and cubic splines finding the same qualitative behaviour using both, which gives some indication of reliability. In the end we chose to use finite differences exclusively since they require less computational time.

\section{Results}
\label{sec: res}

\subsection{Black hole evolution}

\begin{figure}[ht]
\centering
\subfigure[tanh initial data]{
\includegraphics[scale=0.3]{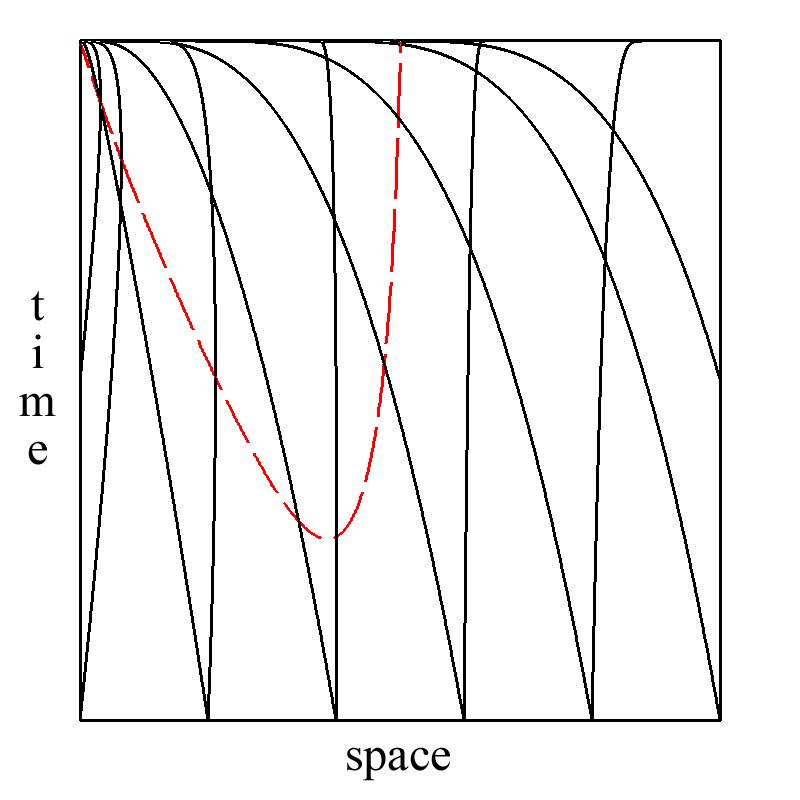}
\label{nullmap tanh}
}
\hspace{0.5in}
\subfigure[gaussian initial data]{
\includegraphics[scale=0.3]{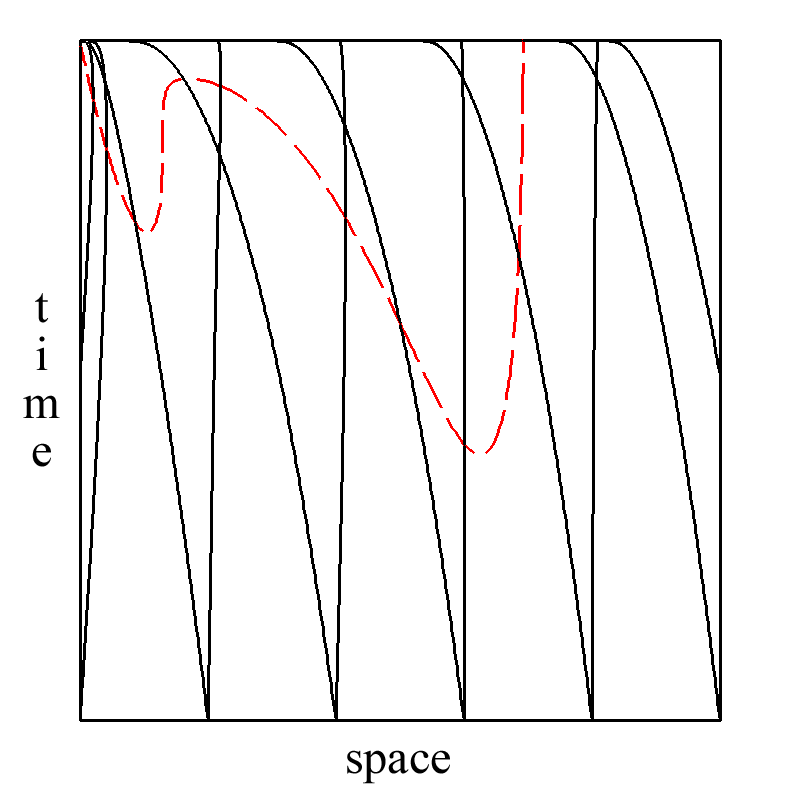}
\label{nullmap gaussian}
}
\caption{Spacetime diagrams of the collapsing massless scalar field using radius for the space axis and PG time for the time axis. The solid black lines are null geodesics and the dashed red lines indicate the trapping surface boundary.}
\end{figure}

Since our coordinates are regular across apparent horizons, we are able to evolve well beyond initial black hole formation up to the point when a singularity begins to form and our boundary conditions at the origin are no longer valid. Spacetime diagrams of black hole formation and evolution are given in figures \ref{nullmap tanh} and \ref{nullmap gaussian} for tanh and gaussian initial data respectively. Using this same data, we have also constructed animations of the mass density profile throughout this process, available for viewing at {\it http://theoryx5.uwinnipeg.ca/users/jziprick/}.

We note that apparent horizons form in pairs within constant PG time slices. This is to be expected from looking at the apparent horizon condition:
\begin{eqnarray}
|\Delta \phi_h|^2 &=& 0,\\
\label{horizon}
\Rightarrow r_h - 2 G^{(4)}{\cal M} &=&0,
\end{eqnarray}
which is initially positive across the $r$-axis. Minima begin to form as mass collapses inward. Black hole formation is indicated by the appearance of a single horizon where the function first dips to zero. This horizon splits into two as the minimum drops below zero; the outer horizon continues to grow as mass passes through it while the inner horizon moves toward the origin. Depending upon the initial mass distribution, subsequent horizon pairs may form as demonstrated in figure \ref{nullmap gaussian} which depicts the evolution of gaussian initial data. The horizon paths define a two dimensional trapping surface within which all null geodesics point towards the origin.

\subsection{Critical behaviour}

In our study of the critical behaviour, we varied the parameters $A$, $B$, and $r_0$ in the gaussian form of initial data, and $A$ in the tanh form. The critical values were found to an accuracy of $\partial p/p \sim 10^{-15}$ using a binary search. We tried a variety of grid resolutions with $r$-spacings near the origin ranging from $10^{-5}$ to $10^{-3}$, which in turn affects the $t$-spacings. The accuracy was roughly equal in all cases where black hole size was not significantly affected by the resolution (when $r_h>>\Delta r(r_h)$). However, finer grids allowed for data to be found for lower values of $ln(p-p_*)$ thus showing critical scaling over a wider range of a given parameter.

To demonstrate scale echoing, we present a plot of the scalar field at $r=0$ as a function of $\tau = ln(t_* - t)$ for the nearest solution to criticality that was numerically achieved using an $r$-spacing near the origin of $10^{-5}$. $t_*$ is the time of black hole formation in the critical solution given by the value approached asymptotically in the small mass limit. Since we choose $\sigma=1$ as our boundary condition at $r=0$, $t$ is equivalent to the proper time at the origin. $\psi(r=0)$ is seen to be periodic in $\tau$ with a period of $\Delta = 3.43 \pm 0.06$ found by averaging over the two oscillations between the minima at $\tau\sim-2.5$ and $-9.3$, with an error determined by the uncertainty in $t_*$. This result is in good agreement with the accepted value of $\Delta \simeq 3.44$.

\begin{figure}
\begin{center}
\includegraphics[width=0.75\linewidth]{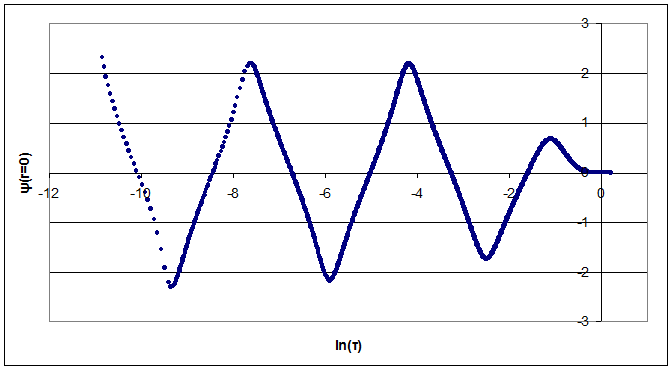}
\caption{A plot of the scalar field at the origin versus $\tau$ illustrating scale echoing in the critical solution. The $r$-spacing in generating this data was $10^{-5}$ near the origin.}
\label{psi0}
\end{center}
\end{figure}

Table \ref{periods} lists the critical exponent and period found for various lattice resolutions and families of initial data using both the mass and maximum curvature scaling relations. Our results in all cases are seen to agree with the accepted values of $\gamma \simeq 0.374$ and $T \simeq 4.6$.

\begin{table}[htb]
\begin{center}
\begin{tabular}{|m{0.75in}|m{0.75in}|m{0.75in}|m{0.75in}|c|c|}
%\begin{tabular}{|c|c|c|c|c|}
\hline
\centering type of scaling & \centering form of initial data & \centering parameter varied & \centering resolution near origin & $\gamma$ & $T$\\
\hline
%\vspace{0.063in}
\centering curvature & \centering gaussian &\centering $A$ &\centering $10^{-5}$ & $0.375 \pm 0.005$ & $4.61 \pm 0.05$\\
\centering mass & \centering gaussian &\centering $A$ &\centering $10^{-5}$ & $0.375 \pm 0.003$ & $4.6 \pm 0.1$\\
\centering mass & \centering gaussian &\centering $A$ &\centering $10^{-4}$ & $0.374 \pm 0.003$ & $4.4 \pm 0.2$\\
\centering mass & \centering gaussian &\centering $B$ &\centering $10^{-4}$ & $0.376 \pm 0.006$ & $4.4 \pm 0.2$\\
\centering mass & \centering gaussian &\centering $r_0$ &\centering $10^{-4}$ & $0.377 \pm 0.005$ & $4.4 \pm 0.2$\\
\centering mass & \centering tanh &\centering $A$ &\centering $10^{-4}$ & $0.371 \pm 0.004$ & $4.5 \pm 0.1$\\
\hline
\end{tabular}
\end{center}
\caption{The critical exponent and period of the mass scaling curve for various initial data and mesh resolutions.}
\label{periods}
\end{table}

\begin{figure}[ht]
\centering
\subfigure[The maximum curvature scaling relation for a variation in the paramter $A$ in gaussian initial data with $\Delta r=10^{-5}$ near the origin.]{
\includegraphics[scale=0.4]{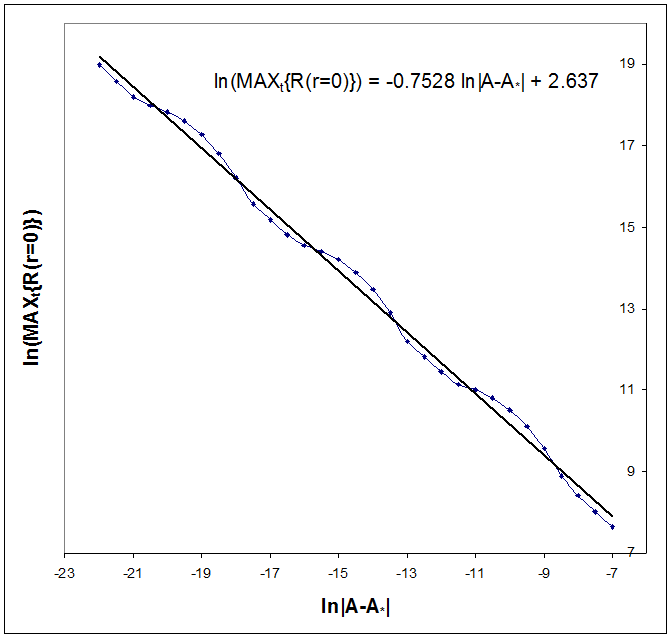}
\label{Ricci plot}
}
\hspace{0.25in}
\subfigure[The mass scaling relation for a variation in the paramter $A$ in gaussian initial data with $\Delta r=10^{-4}$ near the origin.]{
\includegraphics[scale=0.4]{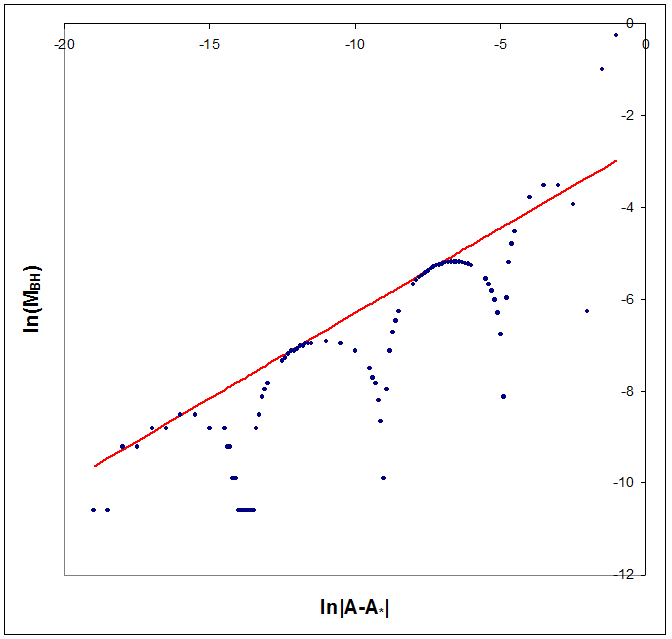}
\label{gaussian plot 1e-4}
}

\vspace{0.5in}

\subfigure[The mass scaling relation for a variation in the paramter $A$ in gaussian initial data with $\Delta r=10^{-5}$ near the origin.]{
\includegraphics[scale=0.4]{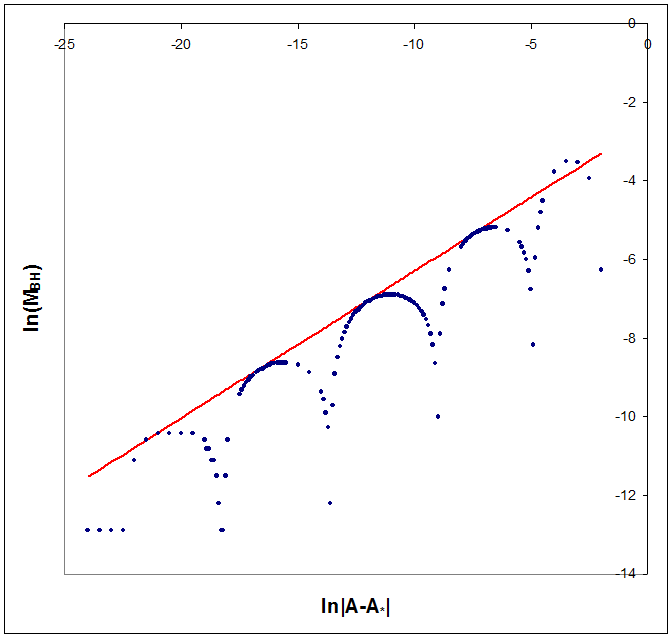}
\label{gaussian plot 1e-5}
}
\hspace{0.25in}
\subfigure[The mass scaling relation for a variation in the paramter $A$ in tanh initial data with $\Delta r=10^{-4}$ near the origin.]{
\includegraphics[scale=0.4]{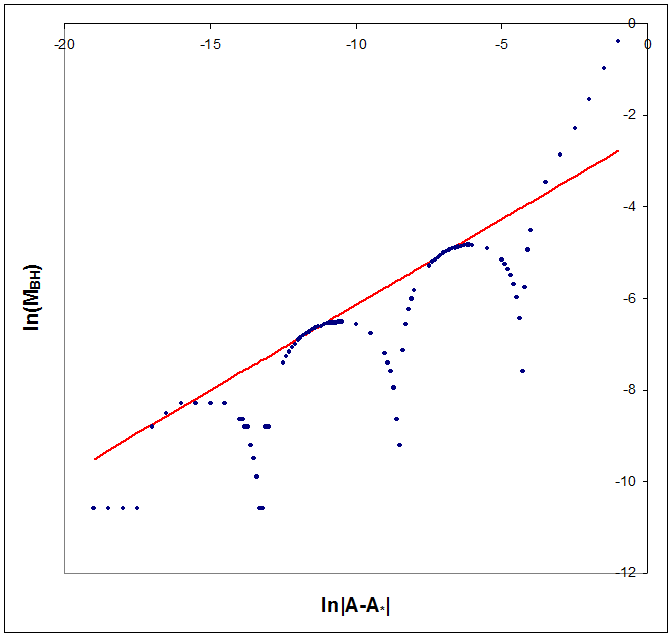}
\label{tanh plot}
}
\caption{Plots of the critical scaling behaviour.}
\label{mass scaling plots}
\end{figure}

A plot of the maximum subcritical curvature scaling behaviour is given in figure \ref{Ricci plot}. $\gamma$ was determined using a least squares linear fit over the points where the solution is seen to follow the intermediate attractor. The error quoted indicates how this value changes depending on the range of points chosen. The period and associated error (given in the first row of table \ref{periods}) were determined by fitting a sine function to the residual left after subtracting the linear fit from this data.

Plots of the supercritical mass scaling behaviour are presented in figures \ref{gaussian plot 1e-4}--\ref{tanh plot}. On each of these we have drawn a line that osculates the peaks of the curves. The values of $\gamma$ were determined by using a least squares linear fit through the data points that are very near this line (within $\sim$ 5~\% of $M_{BH}$)\footnote{Note that our results are insensitive to this choice at the given level of precision.}. The error quoted is twice the standard deviation in the calculated slope. The periods are found by averaging the distance between the cusp-like valleys, with the error representing the uncertainty in locating the cusps. The data used in these calcultions consisted of the two periods corresponding to the lowest range of the parameter that produced black holes such that the size was not significantly affected by lattice resolution (for example, the periods between $ln|A-A_*|\sim-9$ and $-18$ in figure \ref{gaussian plot 1e-5} were used to calculate the values given in the second row of table \ref{periods}).

Previous studies of critical collapse (done in either Schwarzschild or null coordinates) have shown both scaling relationships to be a slight oscillation about a straight line that is well approximated by a sine function. While this is confirmed by our results for the subcritical maximum curvature scaling, it is clearly not the case for our supercritical mass scaling results where we find a periodic form with relatively large amplitude and a distinctly different functional form. Considering that our results agree with the known parameters of Choptuik scaling for varied initial data and mesh resolutions, and that our new-found periodic form is universal within the class of coordinate systems that we are using, it is unlikely that this periodic form is a numerical artifact.

According to Gundlach \cite{Gundlach1} the function $f$ that determines the small wiggle is universal. However, the location and time of formation of the initial apparent horizon can depend on the choice of spacetime slicing. Ours is the first calculation in PG coordinates. These are non-static coordinates for which data are specified on spacelike surfaces that are regular across future horizons. These unique features of PG coordinates are undoubtedly at the root of the new periodic form in the mass scaling relationship. It would be of great interest to continue the near critical evolution past initial horizon formation in order to study explicitly the scaling relation of the final horizon radius, which is independent of spacetime slicing. This is currently under investigation.

\section{Conclusion}
\label{sec: conc}

We have successfully demonstrated Choptuik scaling for a massless scalar field in PG coordinates and revealed an interesting large amplitude component to the oscillations in the mass scaling relation. Our gauge choice allows integration of the dynamical equations beyond horizon formation until the onset of singularity formation. This allowed us to study in detail the evolution of black holes showing explicitly that apparent horizons form in pairs with the number of pairs depending on the complexity of initial data.

One possible direction for work with this system would be to take the number of dimensions as a parameter and compare the results with the dimensional analyses done in \cite{Bland} and \cite{GCD}.

Particularly exciting is the prospect of incorporating quantum corrections into the gravitational potential\footnote{GK is grateful to R.B. Mann and A. Buchel for raising the question about the possible effects of quantum gravity corrections on critical collapse.}.  Loop Quantum Gravity suggests that the underlying discreteness at the quantum level will give rise in the semiclassical limit to an effective short range repulsive component to the gravitational potential which in turn can lead to singularity avoidance \cite{Ashtekar05, bh singularity avoidance}. Husain \cite{Husain08} has done an investigation of the effect of such a correction on Choptuik scaling using double null coordinates which did not allow the code to be extended past horizon formation. Since we are able to investigate the dynamics across the entire spacetime we can investigate directly the effects of short range modifications to the gravitational potential near the onset of singularity formation. We have begun such an investigation \cite{KZ} and preliminary results indicate dynamical singularity avoidance that allows us to calculate numerically the non-singular spacetime showing black hole formation and evaporation. These results will be presented in a future publication. 
 
\section*{Acknowledgements}

We thank Ramin Daghigh, David Garfinkle, Viqar Husain, Randy Kobes, Jorma Louko and Ari Peltola for helpful discussions and correspondence. The authors are grateful to Westgrid for providing the computer resources used for some of the simulations. GK thanks H. Maeda for useful discussions regarding trapping surfaces. This work was supported in part by the Natural Sciences and Engineering Research Council of Canada.

\end{document}